\documentclass[prl,twocolumn,superscriptaddress]{revtex4-2}
\usepackage{amsmath,amssymb,amsfonts,mathrsfs} 	
\usepackage{graphicx}
\usepackage{subfigure}
\usepackage{xcolor}
\usepackage[normalem]{ulem}

\renewcommand{\[}{\begin{equation}}
\renewcommand{\]}{\end{equation}}
\def\bea{\begin{eqnarray}}
\def\eea{\end{eqnarray}}
\def\nn{\nonumber\\}
\newcommand{\intk}{\int_{\rm BZ}  \frac{d{\bf k}}{(2\pi)^d} \;}

\newcommand{\ei}[1]{{\rm e}^{i #1}}
\newcommand{\B}{{\bf B}}

\newcommand{\p}{{\bf p}}

\renewcommand{\k}{{\bf k}}
\newcommand{\vc}{V_{\rm cell}}

\newcommand{\kk}{{\mbox{\boldmath$\kappa$}}}
\def\EEE{\mbox{\boldmath${\cal E}$}}
\def\EE{{\cal E}}

\renewcommand{\v}{{\bf v}}
\renewcommand{\r}{{\bf r}}

\renewcommand{\P}{{\bf P}}
\newcommand{\da}{\partial_{k_\alpha}}
\newcommand{\db}{\partial_{k_\beta}}
\newcommand{\dg}{\partial_{k_\gamma}}

\newcommand{\equ}[1]{Eq.~(\ref{#1})}
\newcommand{\eqs}[2]{Eqs.~(\ref{#1}) and (\ref{#2})}

\def\ket#1{\vert#1\rangle}
\def\ev#1{\langle#1\rangle}
\def\me#1#2#3{\langle#1| \, #2 \, |#3\rangle}
\def\runtime{(\the\time)\qquad\the\month/\the\day/\the\year}
\def\today
 {\count10=\year\advance\count10 by -2000 \number\day--\ifcase
  \month \or Jan\or Feb\or Mar\or Apr\or May\or Jun\or
             Jul\or Aug\or Sep\or Oct\or Nov\or Dec\fi--\number\count10}

\def\hour{\count10=\time\count11=\count10
\divide\count10 by 60 \count12=\count10
\multiply\count12 by 60 \advance\count11 by -\count12\count12=0
\number\count10 :\ifnum\count11 < 10 \number\count12\fi\number\count11}
\begin{document}

\title{Intrinsic nonlinear Hall effect beyond Bloch geometry}

\author{Raffaele Resta}
\email{raffaele.resta.47@gmail.com}
\affiliation{CNR-IOM Istituto Officina dei Materiali, Strada Costiera 11, 34149 Trieste, Italy}
\affiliation{Donostia International Physics Center, 20018 San Sebasti{\'a}n, Spain}

\date{\today}

\begin{abstract} 

The theory of the intrinsic Hall effect, both linear and nonlinear, is rooted in a geometry which is  defined in the Bloch-vector parameter space;  the formal expressions are mostly derived from semiclassical concepts. When disorder and interaction are considered there is no Bloch vector to speak of; one needs a more general quantum geometry, defined in a different parameter space. The
nonlinear Hall effect is a fundamental geometric response of the many-body ground state, not a band-structure peculiarity. The higher-level geometrical formulation of  the intrinsic Hall effect provides very compact expressions, which
have the additional virtue---in the Bloch special case---of yielding the known results in a straightforward way: the logic is not concealed by the algebra.

\end{abstract}

\date{run through \LaTeX\ on \today\ at \hour}


\maketitle 



{\it Introduction.}--The second-order dc conductivity tensor is comprised of three terms which---in a semiclassical treatment---scale as the zeroth, first, and second power of the transport lifetime $\tau$. The first two terms are of the Hall kind, i.e. the induced current is normal to the electric field, and their main entries are quantum-geometrical entities: Berry connections and curvatures in Bloch space \cite{Gao14,Sodemann15,Ma19,Ortix21,Tsirkin22}.
The main focus of the present work is on the $\tau^0$ term, which---to the best of the author's knowledge---was first discovered in  Ref. \cite{Gao14}, where it was attributed to the field-induced ``positional shift" of Bloch electrons; it will be indicated as $\sigma^{(\rm ps)}_{\alpha\beta\gamma}$ in the following. This term has received much attention in recent times \cite{Wang21,Liu21,Das23,Zhang24,Liu24,Chen24}; it is shown here that it is not a peculiarity of Bloch electrons: it is a fundamental geometric response of the many-body ground state.

I adopt here an exact quantum-mechanical framework, where  by ``exact'' I mean that I address a nonrelativistic system of interacting electrons and static classical nuclei, possibly in a disordered configuration. No relaxation time $\tau$ can enter the theory: in fact the concept itself does not make sense beyond the semiclassical theory. 
In the present framework the response functions are causal but nondissipative: therefore the induced current grows with some powers of time. At first order the intrinsic Hall current is $t^0$, and the Drude current is $t^1$ (the electrons undergo free acceleration). At second order one expects three terms, namely $t^0$, $t^1$, and $t^2$, in one-to-one correspondence to the three semiclassical terms, having the same symmetry selection rules: the  $t^0$ and $t^2$ terms require breaking of 
 time-reversal (T) symmetry, besides breaking of  inversion symmetry \cite{Tsirkin22}.  While for the $t^1$ and $t^2$ terms a many-body formulation exists \cite{Watanabe20,rap163}, the $t^0$ term remained elusive so far. Its expression is provided here; for the sake of completeness the other terms are also addressed. 

The parameter space of conventional quantum geometry is  defined by the Bloch vector $\k$, and
 the state vectors are the cell-periodic $\k$-dependent Bloch orbitals; the observables are Fermi-volume integrals. Here instead I adopt the  many-body formulation of quantum geometry \cite{Niu84,Ortiz94,Xiao10,rap165,Onishi25}, where the role of the Bloch vector is played instead by  the ``flux'' $\kk$ entering the kinetic term of the many-body Hamiltonian.  The main entry of $\sigma^{(\rm ps)}_{\alpha\beta\gamma}$ is cast here as a positional-shift tensor: a Berry curvature of an hybrid kind, whose variables are the flux and the electric field. Geometrical tensors of this kind may appear exotic in the context of conductivities; instead they are the main entries  in the theory of polarization (formerly called ``modern'') \cite{Vanderbilt} and of other observables \cite{rap171}.

Many-body quantum geometry addresses  in principle even systems with disorder and correlation,  but has the additional virtue that it allows  for compact and very transparent notations; all geometrical  formula\ae\ can be easily converted when needed---e.g. for DFT implementation---in their more prolix Bloch counterparts. This is shown here in detail for the three terms of second order conductivity.


{\it Exact theory}.--The flux-dependent many body Hamiltonian has been introduced in 1964 by Kohn  \cite{Kohn64}; the most general nonrelativistic Hamiltonian, also accounting for the absence of T-symmetry, reads \[ \hat{H} = \frac{1}{2m} \sum_{i=1}^N \left[\p_i + \frac{e}{c}{\bf A}(\r_i) + \hbar \kk \right]^2 + \hat{V} .  \label{kohn2} \] It  addresses a system of $N$  $d$-dimensional electrons in a cubic box of volume $L^d$. The flux $\kk$ is a vector potential cast in inverse-length units, constant in space; the potential $\hat{V}$ includes the one-body potential (possibly disordered) and electron-electron interaction. 

The system is macroscopically homogeneous; the $\kk$-dependent eigenstates $\ket{\Psi_{n}}$ are normalized to one in the hypercube of volume $L^{Nd}$. 
The thermodynamic limit $N \rightarrow \infty$, $L \rightarrow \infty$, $N/L^d$ constant is understood throughout this work; the $\kk$-derivatives must be evaluated first, and the $L \rightarrow \infty$ limit taken afterwards \cite{Kohn64,Scalapino92,Scalapino93}.

Periodic boundary conditions (PBCs) are assumed: the many-body wavefunctions are periodic with period $L$ over each electron coordinate $\r_i$ independently; the potential $\hat{V}$ enjoys the same periodicity. The vector potential in \equ{kohn2} summarizes all intrinsic T-breaking terms, as e.g. those due to a coupling to a background of local moments; even ${\bf A}(\r)$ obeys PBCs. The vector potential could even account for a macroscopic $\B$ field, provided that it is commensurate, and that the PBCs are modified accordingly \cite{Niu85}.
A time-independent $\kk$ amounts to a pure gauge transformation, but---as shown by Kohn---owing to PBCs the gauge-invariance is broken:  the eigenvalues $E_n$ depend on $\kk$ in the metallic case, while they are $\kk$-independent in insulators  \cite{Kohn64,Scalapino92,Scalapino93,Watanabe18}.

Kohn's Hamiltonian has the virtue that the macroscopic current density can be cast as 
\[ \hat{\bf j} = - \frac{e}{L^d} \hat{\v} , \qquad \hat{\v} =  \frac{1}{\hbar} \partial_{\kk} \hat{H}  \label{current} ;  \] furthermore a flux  linear in time corresponds to perturbing the system with the dc field \[ \EEE   = - \frac{\hbar}{e} \dot\kk . \label{expl} \] 
If $\EEE$  is adiabatically turned on at time $t=0$ the current  response is 
\[  \frac{\partial j_\alpha(t)}{\partial \EE_\beta} =  \frac{e^2}{\hbar L^d} \left[ - \Omega(\kappa_\alpha,\kappa_\beta)  + \frac{t}{\hbar} \frac{\partial^2 E_0}{\partial \kappa_\alpha \partial \kappa_\beta}  \right]  , \label{linear} \] 
where $ \Omega(\kappa_\alpha,\kappa_\beta) = -2 \, \mbox{Im } \ev{\da \Psi_0 | \db \Psi_0}$ is the many-body Berry curvature \cite{Xiao10}. 

The linear conductivity obtains by setting $\kk = 0$:  the two terms in \equ{linear} are $t^0$ (transverse) and $t^1$ (longitudinal), respectively. The longitudinal term in \equ{linear} is rewritten as \[ \partial_t j(t) = \frac{D_{\alpha\beta}}{\pi} \EE_\beta , \quad D_{\alpha\beta} = \frac{\pi e^2}{\hbar^2 L^d} \frac{\partial^2 E_0}{\partial \kappa_\alpha \partial \kappa_\beta} , \label{mona} \] where $D_{\alpha\beta}$ at $\kk=0$ is known as the Drude weight \cite{Kohn64,Scalapino92,Scalapino93}; the Fourier transform of \equ{mona} is \[ -i \omega j_\alpha(\omega) = \frac{D_{\alpha\beta}}{\pi} \EE_\beta(\omega), \] whose causal inversion yields  \[  j_\alpha(\omega) = \frac{D_{\alpha\beta}}{\pi}  \frac{i}{\omega + i\eta}  \EE_\beta(\omega), \quad \eta = 0^+ . \]  The famous Kohn's expression \cite{Kohn64} for the Drude  term in longitudinal conductivity is \[ \sigma^{(\rm Drude)}_{\alpha\beta}(\omega) = \frac{D_{\alpha\beta}}{\pi}  \frac{i}{\omega + i\eta} = D_{\alpha\beta} \left[ \delta(\omega) +\frac{i}{\pi \omega} \right] . \] 

  Both $\Omega(\kappa_\alpha,\kappa_\beta)$ and  
 $D_{\alpha\beta}$ acquire a time dependence when expanded to the next order in $\kk$: \bea \partial_t \Omega(\kappa_\alpha,\kappa_\beta) &=&  \partial_{\kappa_\gamma} \Omega(\kappa_\alpha,\kappa_\beta) \dot\kappa_\gamma, \nn \partial_t  D_{\alpha\beta} &=& \frac{e^2}{\hbar^2 L^d} \frac{\partial^3 E_0}{\partial \kappa_\alpha \partial \kappa_\beta \partial \kappa_\gamma} \dot\kappa_\gamma. \eea Therefore, owing to \eqs{expl}{linear}, one has 
 \bea  \frac{\partial^2 j_\alpha(t)}{\partial \EE_\beta \partial \EE_ \gamma} &=&  \frac{\partial^2 j_\alpha(0)}{\partial \EE_\beta \partial \EE_ \gamma}  + \frac{e^3 t}{\hbar^2 L^d} \dg \Omega(\kappa_\alpha,\kappa_\beta) \nn &-&  \frac{e^3 t^2}{\hbar^3 L^d} \frac{\partial^3 E_0}{\partial \kappa_\alpha \partial \kappa_\beta \partial \kappa_\gamma}   \label{q2} , \eea where all derivatives are evaluated at $\kk = 0$
 
 The three terms are parsed by their $t$-dependence: $t^0$, $t^1$, anf $t^2$, respectively.
 The $t^2$ term accounts for the quadratic Drude conductivity, and coincides with the expression found by Watanabe and Oshikawa \cite{Watanabe20}; the $t^1$  term is the many-body formulation of the nonlinear Hall effect previously established by the present author \cite{rap163}. The $t^0$ term is the main focus of this work: as said above,  it will be indicated as $\sigma^{(\rm ps)}_{\alpha\beta\gamma}$.
 
In order to address this  term it is expedient to make the Hamiltonian $\EEE$-dependent in the scalar-potential gauge, i.e. \[ \hat{V} \rightarrow \hat{V}_0 + e \EEE \cdot \hat{\r} , \qquad  \hat{\r} = \sum_{i=1}^N \r_i ; \] notice that $\kk$ and $\EEE$ are independent variables.
The $\hat{\r}$ operator does not make sense within PBCs, but its offdiagonal elements are {\it defined} as \[ \me{\Psi_0}{\hat{r}_\alpha}{\Psi_n} = i \ev{\Psi_0 | \da \Psi_n} = 
- i \hbar \frac{  \me{\Psi_0}{\hat{v}_\alpha}{\Psi_n}}{{E_0 - E_n}} . \]

The $t^0$ current obtains from the $\EEE$-derivative of the first term in \equ{linear}. Incidentally this is consistent with some alternative nomenclature \cite{Tsirkin22,Liu24}: ``Berry-curvature polarizability''. The concept of field-induced positional shift---within the present beyond-Bloch geometry---can be understood as follows. The electronic term in the polarization of an insulator can be expressed as \[ \P^{(\rm el)} = - \frac{ie}{L^d} \ev{\Psi_0 | \partial_{\kk} \Psi_0}, \quad \kk=0 , \] augmented with a prescription for fixing the gauge \cite{rap162}; the ``position'' of the electrons---i.e. the Berry connection ${\cal A}_\alpha = i \ev{\Psi_0 | \da \Psi_0}$---is ill defined unless the gauge is fixed, yet the ``shift''---i.e. its $\EEE$-derivative---is gauge invariant and well defined, both in insulators and in metals.

In order to make contact with the existing literature I define  the  positional-shift tensor as the derivative with respect to $e\EEE$, i.e. \bea {\cal G}_{\alpha\gamma}  &=& \frac{1}{e} \partial_{\EE\gamma} {\cal A}_\alpha = \frac{i}{e}  \ev{  \partial_{\EE\gamma} \Psi_0 |  \da \Psi_0} + \mbox{c.c.} \nn &=&
  \frac{i}{e}  (\, \ev{ \partial_{\EE\gamma} \Psi_0 |  \da \Psi_0} - \ev{ \da \Psi_0 |    \partial_{\EE\gamma} \Psi_0} \, ) \nn &=& \frac{1}{e} \Omega(\EE_\gamma,\kappa_\alpha) . \label{curva} \eea  ${\cal G}_{\alpha\gamma}$ is therefore cast as an hybrid geometrical tensor of the same family as those entering polarization theory \cite{Vanderbilt}. In fact it is shown elsewhere \cite{rap171} that  the curvature $\Omega(\EE_\gamma,\kappa_\alpha)$, evaluated at $\kk = 0$, yields the linear polarizability of an insulator:  \[  \chi_{\alpha\gamma} =  - \frac{e}{L^d}\Omega(\EE_\gamma,\kappa_\alpha)  . \] Notice that $\Omega(\EE_\gamma,\kappa_\alpha) $ is antisymmetric for the exchange $\EE_\gamma \leftrightarrow \kappa_\alpha$, and is symmetric for the exchange   $\gamma \leftrightarrow \alpha$. 

In the metallic case---considered here---$ \chi_{\alpha\gamma}$ diverges; yet ${\cal G}_{\alpha\gamma}$ retains a physical meaning. In fact the imaginary part of the longitudinal linear conductivity in a metal is \[ \mbox{Im } \sigma_{\alpha\gamma}(\omega) = \frac{D_{\alpha\gamma} }{\pi \omega} + \mbox{Im } \sigma^{(\rm regular)}_{\alpha\gamma}(\omega), \] where the regular term is linear in $\omega$; in insulators the expression is the same but with a vanishing Drude term. Is it then easy to show that  in both insulators and metals one has \[  \lim_{\omega \rightarrow 0} \frac{\mbox{Im }  \sigma^{(\rm regular)}_{\alpha\gamma}(\omega)}{\omega} = \frac{e}{L^d}\Omega(\EE_\gamma,\kappa_\alpha) \label{bingo} . \] In metals this regular term is not observable because it is obliterated by the divergent imaginary part of $\sigma_{\alpha\beta}^{(\rm Drude)}(\omega)$; it manifests itself at second order via its $\kk$-derivatives.

By evaluating ${\cal G}_{\alpha\gamma}$ in the parallel-transport gauge one gets (again both in metals and insulators) \bea {\cal G}_{\alpha\gamma}  &=& - 2  \, \mbox{Re } \sum_{n \neq 0} \frac{\me{\Psi_0}{\hat{r}_\alpha}{\Psi_n}\me{\Psi_n}{\hat{r}_\gamma}{\Psi_0}}{E_0 - E_n} \nn &=& - 2  \hbar^2 \, \mbox{Re } \sum_{n \neq 0} \frac{\me{\Psi_0}{\hat{v}_\alpha}{\Psi_n}\me{\Psi_n}{\hat{v}_\gamma}{\Psi_0}}{(E_0 - E_n)^3} .  \label{rap2} \eea 

Finally, the sought for expression for the positional-shift conductivity is 
\[ \sigma^{\rm (ps)}_{\alpha\beta\gamma}   =  \frac{e^3}{ \hbar L^d}   ( \,\da {\cal G}_{\beta\gamma} - \db {\cal G}_{\alpha\gamma} \, ) ; \label{rap1} \] even  this expression is evaluated at $\kk = 0$.
 \eqs{rap2}{rap1} are   reminiscent of the corresponding Bloch expressions in the  literature:  \eqs{rap3}{rap4} below.

{\it Kohn-Sham theory}.--In the special case of a crystalline system of noninteracting electrons the $L \rightarrow \infty$ limit can be performed analytically thanks to translational symmetry: all intensive observables are then expressed as Fermi-volume integrals.

In a Kohn-Sham framework the ground-state $\ket{\Psi_0}$ is a Slater determinant of single-particle orbitals. Whenever an operator $\hat{O}$ is the sum of one-body operators $\tilde{O}$ its expectation value is given by the sum of the expectation values of $\tilde{O}$ over the occupied orbitals. Here the orbitals are the
Bloch orbitals $\ket{\psi_{j\k}} = \ei{\k \cdot \r} \ket{u_{j\k}}$ with eigenvalues $\epsilon_{j\k}$, normalized in the crystal cell of volume $\vc$---they are normalized differently from $\ket{\Psi_0}$. The formul\ae\ below are  given  per spin channel; for spinless electrons in jargon.

Quantum geometry deals with the $\ket{u_{j\k}}$, which are eigenstates of the Kohn-Sham $\kk$-dependent Hamiltonian  \[ H_\k = \frac{1}{2m} \left( {\bf p} +\frac{e}{c} {\bf A}(\r)  + \hbar \k  + \hbar \kk \right)^2 + V(\r) , \]  hence a $\kk$-derivative evaluated at $\kk = 0$ coincides with a $\k$-derivative. The one-body version of the curvature of \equ{curva} is  then \[ G_{j\alpha\gamma} =  \frac{i}{e} ( \,   \ev{  \partial_{\EE\gamma} u_{j\k} | \partial_{k_\alpha} u_{j\k}}   -\ev{ \partial_{k_\alpha} u_{j\k}  |  \partial_{\EE\gamma} u_{j\k}} \,  ), \label{metric2} \] and the positional-shift conductivity is 
\[ \sigma^{\rm (ps)}_{\alpha\beta\gamma}   =  \frac{e^3}{ \hbar \vc} \sum_{j\k} \theta(\epsilon_{\rm F} - \epsilon_{j\k})  ( \, \partial_{k_\alpha}  G_{j\beta\gamma} -  \partial_{k_\beta}  G_{j\alpha\gamma} \, ) , \label{rapx} \] where $\epsilon_{\rm F}$ is the Fermi energy. In the $L \rightarrow \infty$ limit \[  \frac{1}{ \vc} \sum_{\k} \rightarrow  \intk ,\]
\[ \sigma^{\rm (ps)}_{\alpha\beta\gamma}   =  \frac{e^3}{ \hbar} \sum_{j} \intk \theta(\epsilon_{\rm F} - \epsilon_{j\k})  ( \, \partial_{k_\alpha}  G_{j\beta\gamma} -  \partial_{k_\beta}  G_{j\alpha\gamma} \, ) . \label{rap3} \]
This is identical to the expressions in the semiclassical literature \cite{Gao14,Liu21,Tsirkin22}, once  $G_{j\alpha\gamma}$ evaluated in its equivalent sum-over-states form: 
\[ G_{j\alpha\gamma}  = - 2  \hbar \, \mbox{Re}  \sum_{j' \neq j}  \frac{\me{u_{j\k}}{v_\alpha}{u_{j'\k}} \me{u_{j'\k}}{v_\gamma}{u_{j\k}}}{(\epsilon_{j\k} - \epsilon_{j'\k})^3} \label{rap4} .\] 

It is worth observing that \equ{metric2} is better suited to computational implementation: in fact modern computer codes implement density-functional perturbation theory \cite{Baroni01}, which directly evaluates $\ket{ \partial_{\EEE} u_{j\k}}$ more efficiently than performing sums over states.

When all three terms are included, the quadratic Hall conductivity in the $\omega$-domain is 
\bea \sigma_{\alpha\beta\gamma}(\omega) &=&   \frac{e^3}{ \hbar} \sum_{j} \intk \theta(\epsilon_{\rm F} - \epsilon_{j\k})  \nn
&\times& \bigg[   \partial_{k_\alpha}  G_{j\beta\gamma} -  \partial_{k_\beta}  G_{j\alpha\gamma} 
\nn &+& \frac{1}{\hbar} \partial_{k_\gamma} \tilde{\Omega}_j(k_\alpha,k_\beta) \frac{i}{\omega + i \eta}
\nn &+& \frac{1}{\hbar^2} \frac{\partial^3 \epsilon_{j\k}}{\partial  k_\alpha \partial k_\beta \partial k_\alpha} \left(  \frac{i}{\omega + i \eta} \right)^2\bigg] , \nonumber \eea where $\tilde{\Omega}_j(k_\alpha,k_\beta)$ is the Berry curvature of band $j$ \cite{Vanderbilt}.

Finally, when the infinitesimal $\eta$ is heuristically replaced with the inverse of a relaxation time $\tau$ one gets the semiclassical result, in the form reported e.g. in Ref. \cite{Tsirkin22}. The zero-temperature Fermi function $\theta(\epsilon_{\rm F} - \epsilon)$ can also be heuristically replaced by its finite-temperature counterpart.

{\it Discussion}.---I have shown that both terms of the intrinsic Hall conductivity have a geometrical formulation beyond the Bloch setting, which also applies to a larger class of materials: those where disorder and interaction play an important role. When the system is crystalline the theory is formulated \`a la Kohn-Sham; the resulting expressions coincide with the semiclassical ones in the $\tau \rightarrow \infty$ limit and for zero temperature. This agrees with the common wisdom that the semiclassical approximation is exact when addressing dc transport properties in a crystalline system of non interacting electrons. In this context, the semiclassical approximation is not an approximation, after all.

The major result of this work is the exact many-body expression for  $\sigma^{(\rm ps)}_{\alpha\beta\gamma}$, given as the curl in the $\kk$ variable (the flux) of a positional-shift tensor: a Berry curvature whose variables are the flux and the electric field, having an equivalent  sum-over-states expression. Hybrid geometrical quantities of the same family are at the heart of polarization theory \cite{Vanderbilt}.  This tensor yields indeed the linear static polarizability of a metal, where the divergent Drude term has been discounted.

When the system is cristalline the main entry of $\sigma^{(\rm ps)}_{\alpha\beta\gamma}$ is an analogous band curvature whose variables are instead the Bloch vector and the electric field. The equivalent sum-over-states form coincides with what one finds in the literature  \cite{Gao14,Liu21,Tsirkin22}, while the compact curvature form is possibly new. The latter, as explained above, could be computationally more appealing.

Finally,  a short digression about the extrinsic effects. In the case of the linear Hall conductivity they come in two kinds: $\tau^0$ (called ``side-jump'') and $\tau^1$ (called ``skew scattering'') \cite{Nagaosa10}, while the terms ``intrinsic'' and ``geometrical'' are used as synonymous  (the Bloch geometry of the pristine crystal is intended). When the disordered system is addressed by means of  a supercell (ideally in the $L  \rightarrow \infty$ limit) the effects of disorder become by construction intrinsic; we previously argued in Ref. \cite{rap149} that the geometrical conductivity of the disordered system includes the side-jump contributions, besides the pristine-crystal geometrical response.
A similar statement holds for the  quadratic Hall conductivity: the $\tau^0$ extrinsic effects become intrinsic and included in the present generalized geometrical formulation of $\sigma^{(\rm ps)}_{\alpha\beta\gamma}$.

{\it Acknowledgments.}---I am deeply indebted to Ivo Souza for the many invaluable conversations we had about this topic.
Work supported by the Office of Naval Research (USA) Grant No. N00014-20-1-2847.
 

\end{document}